\documentclass[fleqn,usenatbib]{mnras}

\usepackage{newtxtext,newtxmath}

\usepackage[T1]{fontenc}

\DeclareRobustCommand{\VAN}[3]{#2}
\let\VANthebibliography\thebibliography
\def\thebibliography{\DeclareRobustCommand{\VAN}[3]{##3}\VANthebibliography}

\usepackage{graphicx}	
\usepackage{amsmath}	
\usepackage{parskip}
\usepackage{enumerate}

\title{Search and detection of pulsars in the monitoring observations at the frequency 111 MHz}

\author[S. A. Tyul'bashev et al.]{
S.A. Tyul'bashev,$^{1}$\thanks{E-mail: serg@prao.ru (SAT)}
V.S. Tyul'bashev,$^{1}$
M.A. Kitaeva,$^{1}$
A.I. Chernyshova,$^{1,2}$
V.M. Malofeev,$^{1}$
\newauthor {
I.V. Chashey$,^{1}$
V.I. Shishov$,^{1}$
R.D. Dagkesamanskii$,^{1}$
S.V. Klimenko,$^{3}$
I.N. Nikitin,$^{4}$
L.D. Nikitina$^{4}$}
\\
$^{1}$ P.N. Lebedev Physical Institute of the Russian Academy of Sciences, Astro Space Center, Pushchino Radio Astronomy Observatory,\\
Radiotelescopnaya 1a, Moscow reg., Pushchino, 142290, Russia \\
$^{2}$ Pushchino State Institute of Natural Science\\
$^{3}$ Moscow Institute of Physics and Technology, Institute of Physical and Technical Informatics\\
$^{4}$ Fraunhofer Institute for Integrated Circuits IIS, Germany\\
}

\date{2019}

\pubyear{2019}

\begin{document}
\label{firstpage}
\pagerange{\pageref{firstpage}--\pageref{lastpage}}
\maketitle

\begin{abstract}
Pulsars with periods more than 0.4 seconds in the declination range $-9^{\circ} < \delta < 42^{\circ}$ and in the right ascension range $0^h < \alpha < 24^h$ were searched in parallel with the program of interplanetary scintillations monitoring of a large number of sources with the radio telescope LPA LPI. Four-year observations carried out at the central frequency 110.25 MHz in the band 2.5 MHz and in six frequency channels with the time resolution 100 ms were used for the search. Initial detections of candidates for pulsars were based on the Fourier power spectra averaged over the whole observations period. The candidates for pulsars were checked in more details from observations analysis with higher time-frequency resolution: 32 frequency channels with time resolution of 12.5 ms. 18 new pulsars was discovered, their main characteristics are presented.
\end{abstract}




\section{Introduction}

Recently, a number of surveys have been carried out on the search for new pulsars at the world's largest radio telescopes. Of such observations in regions of the sky larger than the steradian, it should be noted the surveys of $NHTRU$ and $HTRU$, which are conducted at a frequency of $1.4$\,GHz on radio telescopes of $100$ meters and $64$ meters in Effelsberg and Parks (\citeauthor{Barr2013} (\citeyear{Barr2013}), \citeauthor{Keith2010} (\citeyear{Keith2010})), surveys of $GBNCC$ and $AO327$, conducted at frequencies of $350$,MHz and $327$ MHz on radio telescopes of $100$ meters and $300$ meters in Green Bank and Arecibo (\citeauthor{Boyles2013} (\citeyear{Boyles2013}), \citeauthor{Deneva2013} (\citeyear{Deneva2013})). Surveys of LOTAAS and LPPS were conducted with different accumulations and different time and angular resolution at a lower frequency, $142$\,MHz, using the LOFAR (\citeauthor{Coenen2013}, \citeyear{Coenen2013}) aperture synthesis system. These surveys were carried out on the world's best radio telescopes, whose receiving equipment, frequency of observations, bands and time constants are specially selected for search operations in such a way as to ensure maximum sensitivity during pulsar search. Despite the fact that the surveys are conducted in areas that have already been repeatedly viewed, including on the same antennas, nevertheless, it is possible to detect new pulsars, both in new and previously processed archival records.

The detection of new pulsars in previously viewed areas of the sky is associated with several factors. Firstly, digital receivers are used in the new observations, which make it possible to implement very wide frequency bands of observations, and thereby obtain a higher sensitivity than was in the early surveys. Secondly, new methods of processing observations are being developed that allow searching for pulsars by reprocessing archived data (\citeauthor{Keane2010}, \citeyear{Keane2010}). Thirdly, new ways of searching for pulsars are emerging, which allow searching for rare types of transient pulsars (RRATs) (see, for example, \citeauthor{Burke2011} (\citeyear{Burke2011})).

In the Pushchino Radio Astronomy Observatory of the Astrocosmic Center of the Physical Institute of the Russian Academy of Sciences (PRAO ASC LPI) on a Large Phased Array (LPA), after a major reconstruction of the antenna for about 5 years, regular monitoring of compact radio sources scintillating on interplanetary plasma is carried out. Observations are carried out around the clock on digital radiometers, in a total band of 2.5 MHz, in six frequency channels, with a time resolution of $0.1$s, in 96 spatial beams, in a sector overlapping $50^{\circ}$ declination (\citeauthor{Shishov2016},  \citeyear{Shishov2016}). The primary data can be used for a number of tasks, including the search for pulsars.

An early search for pulsars in the PRAO was carried out from the moment the first pulsars were reported, but difficulties associated with the lack of multibeam observations and the slow processing speed of large data arrays made it possible to detect only 6 new pulsars  (\citeauthor{Vitkevich1969} (\citeyear{Vitkevich1969}), \citeauthor{Shitov1980} (\citeyear{Shitov1980}), \citeauthor{Shitov2009} (\citeyear{Shitov2009})). A new search for pulsars is carried out using archived data from the LPA LPI, which since August 2014 have been obtained from simultaneous observations in two time-frequency modes (see the next paragraph and the work of \citeauthor{Tyulbashev2016} (\citeyear{Tyulbashev2016})).

In the last two years, different groups have attempted to search for new pulsars in two different ways in PRAO LPI. The first method is a direct enumeration of periods and dispersion measures. In total, 24 days of observations were processed and 7 new pulsars were found (\citeauthor{Tyulbashev2016}, \citeyear{Tyulbashev2016}). The second method is the use of Fourier power spectra. Three months of observations were processed and 14 pulsar candidates (\citeauthor{Rodin2015}, \citeyear{Rodin2015}) were found. These works show that the LPA LPI antenna is an effective telescope for searching for pulsars. The cardinal advantage of observations on the LPA LPI radio telescope, in contrast to all the surveys listed above, is long-term and regular monitoring, that is, daily observations for several years. In this paper, an attempt is made to use this advantage.

\section{Observations and processing of observations}

LPA LPI is a diffraction array having fixed beam directions of the emission pattern in the sky, operating at a central frequency of 110.25 ~ MHz and having a geometric area of 72000 sq.m. Several independent radio telescopes are currently implemented on the basis of LPA LPI. The first radio telescope has 512 beams of the emission pattern with the intersection of the beams of the diagram at about 0.8. The effective area of this radio telescope is approximately 20,000 sq.m. in the direction to the zenith, and observations can be made on it at declinations of $-20^{\circ}<\delta<88^{\circ}$. This radio telescope is mainly used to study pulsars in single-beam mode on radiometers specialized for pulsar observations. The second radio telescope has 128 beams of the emission pattern with the intersection of the beams of the diagram at 0.4 and with the possibility of observations on declinations $-9^{\circ}<\delta<55^{\circ}$. At the moment there are 96 digital radiometers that are connected to 96 beams of this emission pattern. The effective area of the second radio telescope in the direction of the zenith is about 45,000 sq.m. The third radio telescope is undergoing modernization, after which a 512 beam antenna will be obtained, having an effective area comparable to the area of the second radio telescope, and the ability to observe in several beams simultaneously. The fourth radio telescope is used in test experiments related to the modernization of the antenna. At the same time, observations on scientific programs are not interrupted during tests.

Regular round-the-clock monitoring observations on the second radio telescope of the LPA LPI were started in 2013 after the reconstruction of the antenna. After connecting all the radiometers, they were conducted at declinations of $-9^{\circ}<\delta<42^{\circ}$ in 96 beams of the LPA antenna in six frequency channels with a time resolution of $100$~ms and a total observation band of 2.5 MHz. Since August 2014, a parallel recording of the signal has been carried out in the second time-frequency mode: 32 frequency channels and a time resolution of 12.5 ms. Details concerning the reconstruction of the radio telescope, a new digital radiometer and a specially created program for processing observations are described in detail in the works \citeauthor{Tyulbashev2016} (\citeyear{Tyulbashev2016}), \citeauthor{Shishov2016}  (\citeyear{Shishov2016}).

Monitoring observations can be used in various tasks, and in our previous work (\citeauthor{Tyulbashev2016}, \citeyear{Tyulbashev2016}) they were used to search for new pulsars. The search was done by iterating over periods from 0.5 to 15 seconds and iterating over dispersion measures within $0-200~pc/cm^3$. The criterion for detecting a pulsar was a set of four signs:

\begin{enumerate}[1)]
\item repeatability of the signal by sidereal time (at least three records out of all processed);
\item the presence of a pronounced maximum on the S/N dependence in the average profile on the DM;
\item the presence of a S/N ratio greater than 6 in at least one of the confirming pulsar existence of the original record;
\item on a record with a double period on the dispersion measure expected for this pulsar, the average profiles have approximately the same height.
\end{enumerate}

Obviously, if a pulsar does not have its own or interstellar medium-related flux density variability, and its period and the derivative of the period are known with high accuracy, even from observations at other frequencies, then it is possible to accumulate pulses for all observation sessions and thereby increase the S/N ratio proportional to the square root of the accumulated impulses. In this case, weak pulsars can be observed on the antenna, which cannot be detected during a short-term search. However, if the pulsar has not been observed before or its period is known with low accuracy, then accumulation taking into account the change in the arrival time of the pulsar pulse phase is impossible. Therefore, in the initial search for weak pulsars in monitoring observations, it is necessary to use a technique that allows us to detect candidates for weak pulsars and does not require us to know the pulse arrival phase on a daily basis.

This technique has been used for a long time in the search for pulsars, including with us (\citeauthor{Tyulbashev2000}, \citeyear{Tyulbashev2000}) and consists in searching for harmonics in the power spectrum. We plotted the Fourier power spectrum at time intervals of 2048 points (approximately 3.5 minutes of recording) and, to increase sensitivity, we added the spectra corresponding to one sidereal time for the entire observation period. In the power spectrum, information about the pulse arrival phase is lost, but when the records are added together, the S/N ratio increases for harmonics that are multiples of the pulsar period. Previously, observations with strong interference were discarded. They made up about $25-30\%$ of the total time. If the pulsar flux density does not change from day to day and at the same time the gain of the radiometers changes slightly, then processing four years of observations should lead to an increase in the S/N ratio by about 30 times. Considering this circumstance and using the theoretical estimate of the sensitivity of observations at the LPA LPI at a frequency of 111 MHz, given in \citeauthor{Tyulbashev2016} (\citeyear{Tyulbashev2016}), we can expect that the best sensitivity when searching for pulsars will be equal to 0.2 and 0.6 mJy for pulsars located outside and inside the plane of the Galaxy. Taking into account the practical sensitivity of observations obtained in the same work, the expected sensitivity will be three times worse. Nevertheless, it should be higher than the sensitivity when searching for pulsars in any currently conducted surveys, given that the average spectral index is $(\alpha)$ of the pulsar spectrum in the range of 102-400 MHz is approximately equal to 1.35 ($S\sim\nu^{-\alpha}$) (\citeauthor{Malofeev2000} (\citeyear{Malofeev2000}), \citeauthor{Bilous2016} (\citeyear{Bilous2016})).

The criteria for detecting pulsars used in the previous work \citeauthor{Tyulbashev2016} (\citeyear{Tyulbashev2016}) and given in the paragraph above were supplemented during the search for pulsars by averaged power spectra, taking into account that in the most realistic case, a pulsar is most often absent in an individual record. These criteria are as follows:

\begin{enumerate}[1)]
\item signal repeatability by sidereal time;
\item the presence of a pronounced maximum on the S/N dependence on the dispersion measure (in the analytical part of the pulsar research program, a block of addition of such dependencies was added (see https://github.com/vtyulb/BSA-Analytics ));
\item there must be at least two harmonics in the power spectrum (this condition leads to the fact that the search is carried out for pulsars with periods greater than 400 ms);
\item it is desirable to have at least one record obtained for the 32-channel frequency mode confirming the existence of a pulsar with a S/N ratio observed in the average profile greater than 6. At the same time, on a record with a double period, the average profiles should have approximately the same height.
\end{enumerate}

The appearance of point 3 is caused by the fact that when analyzing the results of processing observations, several hundred single harmonics were revealed that do not repeat on other right ascensions and declinations, and several dozen single harmonics that repeat in sidereal time on different right ascensions and declinations. These results require additional analysis and are not considered in this paper.

\begin{figure}
\begin{center}
	\includegraphics[width=0.4\textwidth]{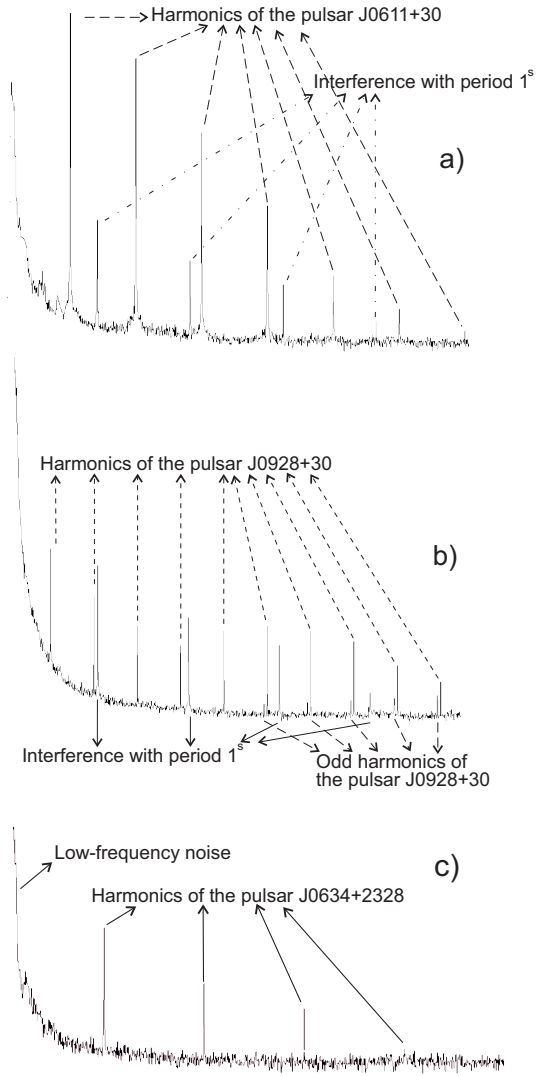}
    \caption{The power spectra of pulsars are $J0611+30$ (7 harmonics are visible), $J0928+30$ (10 harmonics are visible) and $J1634+2328$ (4 harmonics are visible).}
    \label{fig:tul_fig1_E}
\end{center}
\end{figure}

In Fig.~\ref{fig:tul_fig1_E} examples of Fourier power spectra of several pulsars detected during survey are given. Figure 1a shows the little-studied pulsar J0611+30, detected at the LPA LPI with a period of $P = 1.4116$~s. This pulsar was discovered on Arecibo (\citeauthor{Camilo1996}, \citeyear{Camilo1996}). An estimate of its flux density at 430 MHz $S_{430}=1.6$ mJy. Pulsar period ($P_{ATNF} = 1.412090$ s (http://www.atnf.csiro.au /people/pulsar/psrcat/), known only up to the sixth decimal place. In the figure between the strong harmonics, relatively weak harmonics are visible, which are associated with periodic interference observed on the LPA LPI in individual beams of the emission pattern. Figure 1b shows the power spectrum of the pulsar J0928+30. This relatively strong pulsar ($P = 2.0919$~s) was detected at the LPA LPI during a search with a direct search of periods and dispersion measures (\citeauthor{Tyulbashev2016}, \citeyear{Tyulbashev2016}). A complex visible set of harmonics consists of the harmonics of the pulsar itself, its non-multiple harmonics, as well as periodic interference ($P = 0.9981 $ ~ s), which is also visible in Fig.1a. Figure 1c shows the power spectrum of one of the weak new pulsars (J1634+2328). The period of this pulsar is $P = 1.2087$s, and there is no periodic interference in this summed power spectrum.

As mentioned above, the applied method of searching for pulsars will be optimal for pulsars that do not show variability in flux density, or for pulsars that have flux variability, but are strong enough to be regularly observed in individual records. In the section Discussion of the results, the problem of searching for weak pulsars with variable emission will be considered separately.

\begin{table*}
	\centering
	\caption{Upper estimates of the integral flux density of the gamma pulsars}
	\label{tab:tab1}
	\begin{tabular}{ccccccc}
		\hline
name & $\alpha_{2000}$ & $\delta_{2000}$ & P(c) & $DM$(pc/cm$^3$) & $W_{0,5}$(mc) & wp\\
		\hline
J0122+1407 & 01h22m07s & $14^\circ 07^{'}$ & 1.3885 & $18 \pm 2$ & 30 & 2 \\
J0609+1634 & 06h09m13s & $16^\circ 34^{'}$ & 0.9458 & $84 \pm 5$ & 45 & 1 \\
J0810+3725 & 08h10m30s & $37^\circ 25^{'}$ & 1.2483 & $16 \pm 2$ & 35 & 1 \\
J0933+3245 & 09h33m50s & $32^\circ 45^{'}$ & 0.9616 & $18 \pm 3$ & 30 & 1 \\
J1243+1752 & 12h43m00s & $17^\circ 52^{'}$ & 1.2165 & $5 \pm 3$ & 200 & 1 \\
J1528+4053 & 15h28m00s & $40^\circ 53^{'}$ & 0.4764 & $6 \pm 2$ & 110 & 2 \\
J1536+1749 & 15h36m15s & $17^\circ 49^{'}$ & 0.9333 & $28 \pm 3$ & 20 & 3 \\
J1634+2328 & 16h34m50s & $23^\circ 28^{'}$ & 1.2087 & $36 \pm 4$ & 40 & 2 \\
J1637+4003 & 16h37m30s & $40^\circ 03^{'}$ & 0.7675 & $31 \pm 4$ & 35 & 3 \\
J1644+1346 & 16h44m15s & $13^\circ 46^{'}$ & 1.0989 & $34 \pm 1.5$ & 20 & 1 \\
J1651+1422 & 16h51m00s & $14^\circ 22^{'}$ & 0.8280 & $48 \pm 10$ & -- & 3 \\
J1657+3248 & 16h57m30s & $32^\circ 48^{'}$ & 1.5702 & $23 \pm 2$ & 45 & 2 \\
J1832+2758 & 18h32m10s & $27^\circ 58^{'}$ & 0.6318 & $46 \pm 3$ & 30 & 2 \\
J1844+4117 & 18h44m45s & $41^\circ 17^{'}$ & 0.9157 & $50 \pm 10$ & 300 & 3 \\
J1932-0108 & 19h32m15s & $-01^\circ 08^{'}$ & 0.5937 & $35.5 \pm 1.5$ & 25 & 1 \\
J1954+3007 & 19h54m00s & $30^\circ 07^{'}$ & 1.2710 & $42 \pm 2$ & 35 & 2 \\
J2051+1315 & 20h51m30s & $13^\circ 15^{'}$ & 0.5532 & $42.5 \pm 2.5$ & 120 & 2 \\
J2350+3148 & 23h50m10s & $31^\circ 48^{'}$ & 0.5081 & $37.5 \pm 2.5$ & 15 & 1 \\
		\hline
	\end{tabular}
	\label{tab:tab1}
\end{table*}

\section{Results}

After analyzing the processed data, 18 new pulsars were found that meet the criteria listed above. All these objects can be divided into three parts. The first part is strong pulsars, for which it was possible to find such individual records in 32-frequency data that it is possible to construct a dynamic spectrum, the S/N dependence on the dispersion measure and average profiles with a double period from them. The second part are pulsars for which there are records with a S/N dependence on the dispersion measure and average profiles, but it is not possible to obtain a dynamic spectrum from individual records. The third part is the weakest pulsars. For them, it was possible to find individual records in 6-frequency data and no more than one individual record for 32-frequency data. Apparently, for the confident detection of such weak pulsars, we lack the sensitivity of the LPA LPI antenna in individual observation sessions.

Figure 2 shows the new pulsars, which we have divided into two groups according to the observed emission intensity. For the strongest pulsars, the figure shows the average profiles with a double period and dynamic spectra for the best selected observation day. On dynamic spectra, frequencies grow from top to bottom. The spectra are given for a double period. To increase the contrast in some dynamic spectra, smoothing with a moving average was performed. For medium-strength pulsars and weak pulsars, the S/N ratio in individual channels does not allow us to construct dynamic spectra. The figure for these pulsars shows only the average profiles with a double period. For the weakest pulsars, the quality of the average profiles is low due to the poor S/N ratio, and for one pulsar it was not possible to build average profiles based on 32-frequency data. The division of pulsars into strong and medium + weak ones introduced by us is very conditional. Due to the different sensitivity of the LPA LPI antenna in different directions across the sky, the observed flux densities of two identical discrete radio sources, with a difference of their coordinates in declination of several degrees, may differ several times (see \citeauthor{Shishov2016} (\citeyear{Shishov2016})). Therefore, a pulsar, conventionally called weak in our observations, may turn out to be medium or even strong in observations on other radio telescopes.

\begin{figure*}
\begin{center}
	\includegraphics[width=0.8\textwidth]{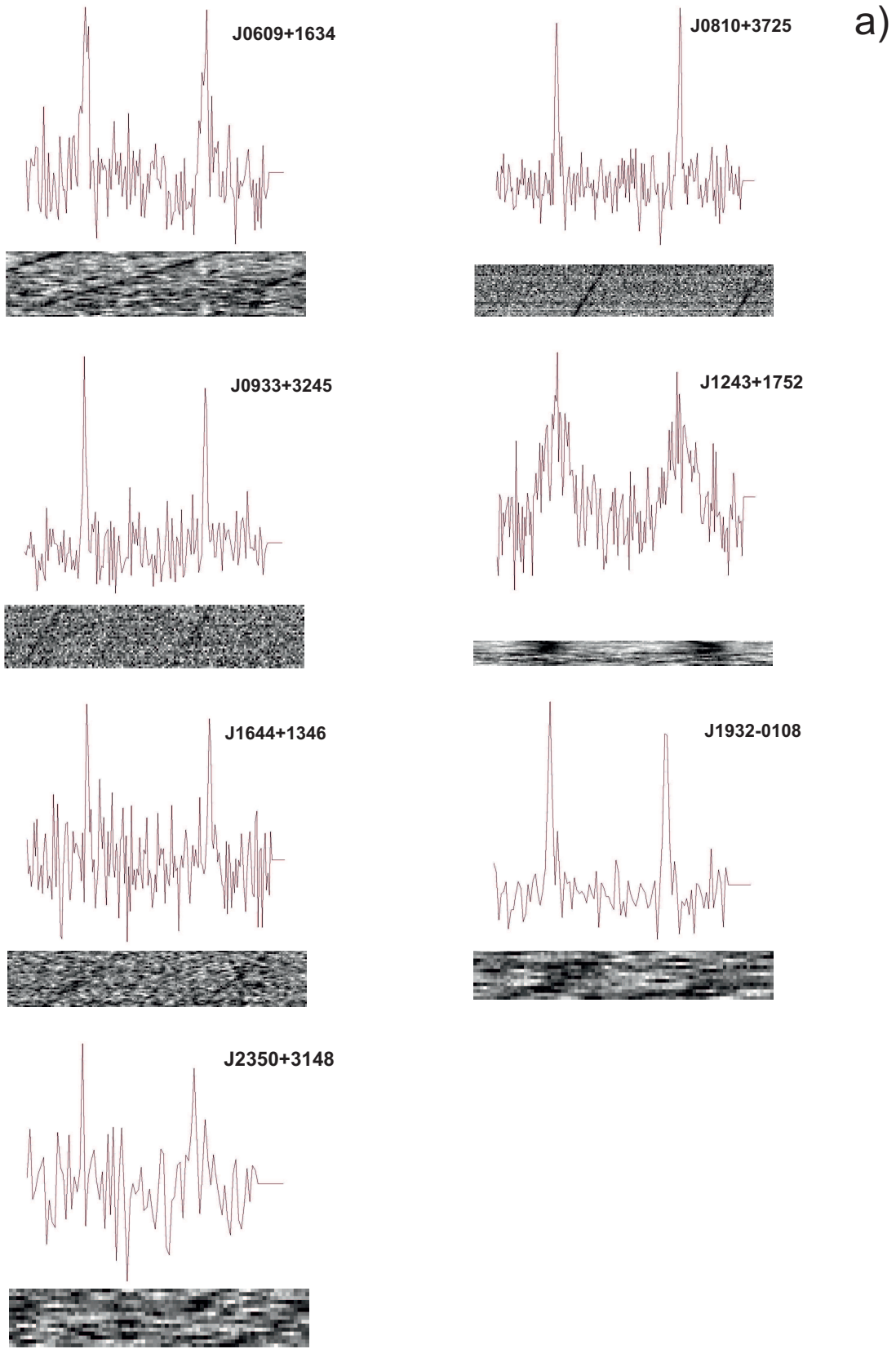}
    \caption{Dynamic spectra and average double-period profiles for detected pulsars.}
    \label{fig:tul_fig2a_E}
\end{center}
\end{figure*}

\begin{figure*}
\begin{center}
	\includegraphics[width=0.8\textwidth]{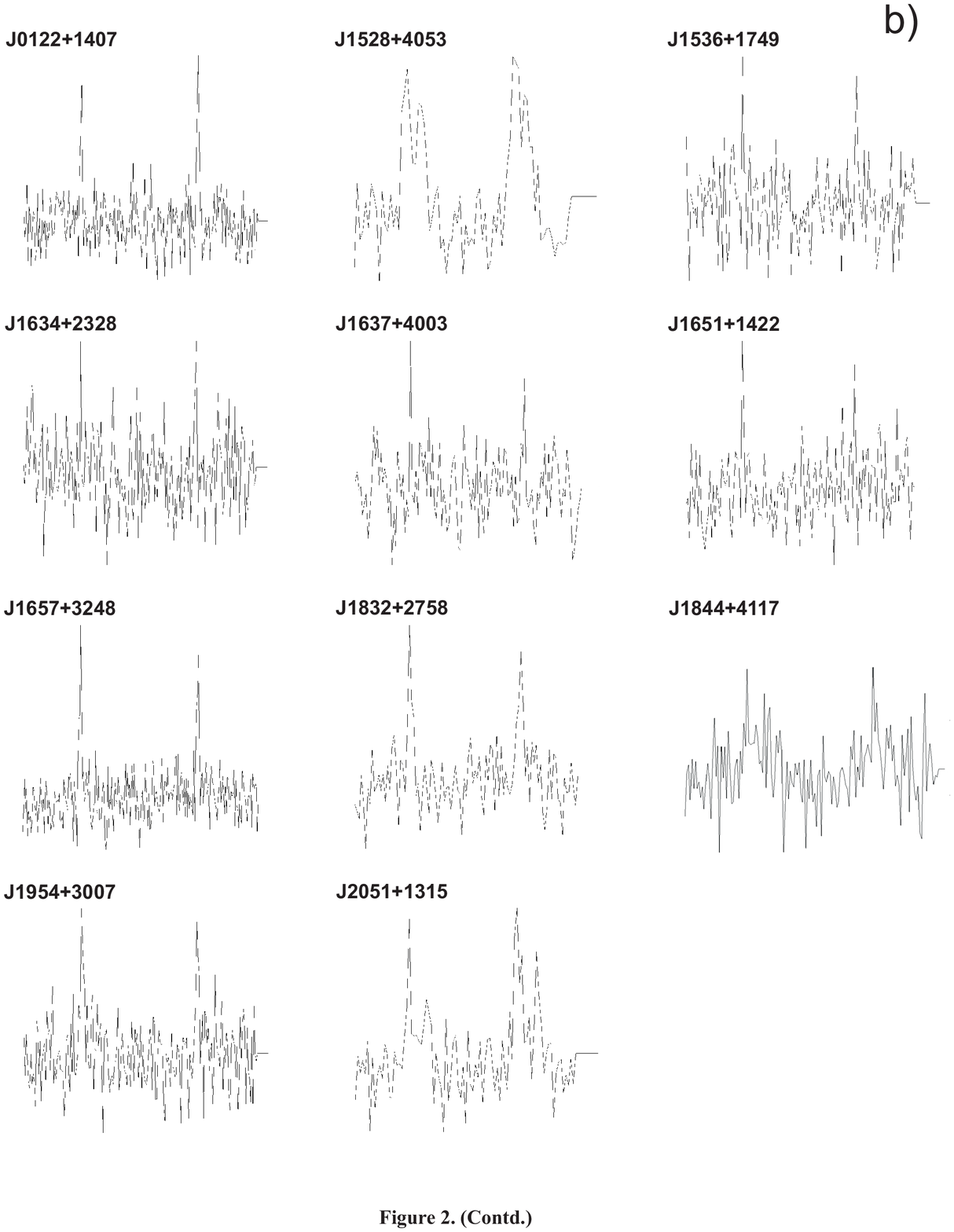}
    \caption{Dynamic spectra and average double-period profiles for detected pulsars.}
    \label{fig:tul_fig2b_E}
\end{center}
\end{figure*}

The Table~\ref{tab:tab1} shows the main characteristics of the new pulsars. The first column of the table shows the name of the pulsar in the annotation J2000. In the second and third columns are the coordinates of the pulsar in direct ascension and declination for the year 2000. The accuracy of the coordinates for the right ascension is $\pm 1^m$, with the exception of the pulsar J1932-0108, which has a coordinate error of $\pm 1.5^m$. By declination, the accuracy of the coordinates is $\pm 20^\prime$, with the exception of the pulsar J1932-0108, which has an error of $\pm 30^{'}$. Columns 4-6 give the pulsar period $(P)$, its DM and the half-width of the average profile ($W_{0.5}$). The error in determining the period is $\pm 0.0005$~s, with the exception of pulsar J1932-0108, which has an expected error of 0.001 s. The S/N ratio for most of the new pulsars was low, so the estimates of $W_{0.5}$ may be greater or less than the ~\ref{tab:tab1} estimates given in the table with a coefficient of 1.5. The column 7 indicates the power of the pulsar. The number one indicates strong pulsars, the number two indicates pulsars of medium strength, the number three is weak pulsars.

Of the 18 pulsars detected, 9 lie in areas with minimal background temperature and are far from the Galactic plane. The median value of the dispersion measure of these pulsars is about $20 pc/cm^3$. 4 pulsars lie close to the plane of the Galaxy. Their median dispersion measure is about 40 $pc/cm^3$. 5 pulsars are located in the area of the north polar hole. The median value of the dispersion measure of these pulsars is about 30 $pc/cm^3$. Separately, we note pulsars, which, in our opinion, may be of particular interest. Pulsars J1243+1754, J1528+4053, J1844+4118, J2051+1332 have wide average profiles and, apparently, are close to coaxial rotators. Pulsar J1243+1754 is located in the region of the north polar hole, where an increased concentration of electrons can be expected, and at the same time its $DM =5 \pm 3 pc/cm^3$, which, taking into account the error in determining the dispersion measure, potentially makes it one of the closest known pulsars.

\section{Discussion of the results}

During the search, 131 pulsars were found, 18 of which were discovered for the first time. Including found 7 pulsars from the previous publication on the search
for pulsars (\citeauthor{Tyulbashev2016}, \citeyear{Tyulbashev2016}). Approximately $1/4$ of known pulsars were found in the area $-9^\circ<\delta<21^\circ$, and approximately $2/3$ of known pulsars were found in the area $21^\circ<\delta<42^\circ$.

There may be several reasons why it was not possible to detect such a large number of known pulsars. The main reasons, in our opinion, are three.

Firstly, due to the fact that the LPA LPI is a diffraction array with a fixed position of the beams of the emission pattern in height, its sensitivity is very different in different directions across the sky. According to the work \citeauthor{Shishov2016} (\citeyear{Shishov2016}), the sensitivity of the LPA LPI is maximal in the direction to the zenith, which corresponds to the declination of $55^\circ$. Based on this direction, all estimates of the maximum sensitivity of the radio telescope were made. From Fig.1 in \citeauthor{Shishov2016} (\citeyear{Shishov2016}), it is easy to see that the sensitivity of the radio telescope at the observed declinations from $+42^\circ$ to $-9^\circ$ generally decreases according to the cosine law and decreases from the level of 0.9 to the level of 0.4 from the maximum sensitivity. Losses due to the envelope of the formed eights of beams reach at the edges the level of 0.9 from the maximum possible for these eights. If the direction to the source is exactly in the middle between the directions of the LPA LPI beams in the configuration used for monitoring, then the sensitivity of the radio telescope is about 0.4 compared to the case when the direction to the source exactly coincides with the direction of the LPA LPI beam. For antenna beams at low declinations, the effective reception band is additionally reduced due to the fact that there is a frequency dependence between the frequency of observations and the beam direction of the antenna pattern. Thus, the real sensitivity of the LPA LPI radio telescope strongly depends on the difference between the direction to the source under study and the beam direction of the emission pattern and may be in the worst case about 10 times less than for the ideal case considered in Table 2 of the work \citeauthor{Tyulbashev2016} (\citeyear{Tyulbashev2016}) on the detection of pulsars at declinations between $+21^\circ$ and $+42^\circ$ in single entries. Since the coordinates of the newly discovered pulsars are known with low accuracy, it is difficult to take into account this effect on the observed pulsar flux density. At the same time, there is no doubt that some of the pulsars will be lost during the search due to the low sensitivity of the LPA LPI in certain directions.

\begin{figure}
\begin{center}
	\includegraphics[width=0.45\textwidth]{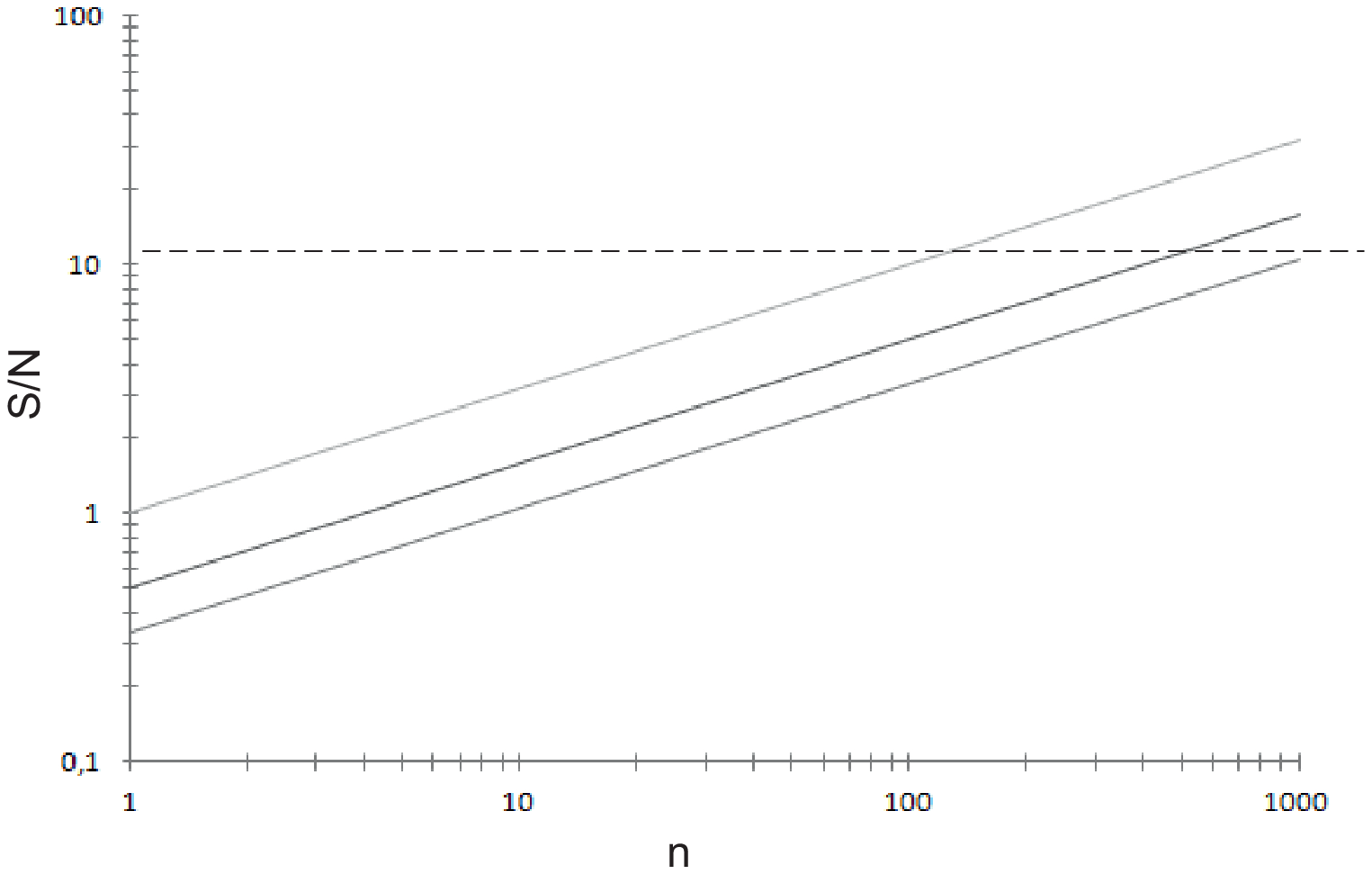}
    \caption{Changing the S/N ratio depending on the initial noise when accumulating over $n$ days}
    \label{fig:tul_fig3_E}
\end{center}
\end{figure}

Secondly, pulsars have variable emission, and when adding up the power spectra, this factor was not taken into account in any way. In the section "Observations and processing of observations", it has already been said that the method used to search for pulsars will be optimal for pulsars that do not have their own variability, and at the same time signal scintillations caused by the interstellar medium in the direction of the pulsar should be significantly suppressed so as not to cause modulation of the flux density. It is known that the interstellar medium for pulsars causes diffraction and refractive scintillation. Diffraction scintillation is fast (from seconds to minutes) and strong (the degree of scintillation can reach {100\%}. Refractive scintillations are slow (from days to years) and weaker. Depending on the frequency of observations, the observation band, the width of the frequency channel and the dispersion measure of pulsar, the observed variability of the flux density due to interstellar scintillation can vary widely.

If we talk about interstellar scintillation observed in the meter wavelength range, then we can see that the variability caused by scintillation can change the observed flux density several times on daily time scales. The variability due to refraction and diffraction is strongest for close pulsars, so there is a clear shortage of close pulsars, noted in \citeauthor{Malofeev2000} (\citeyear{Malofeev2000}). Studies of diffraction scintillation at a frequency of $102$ MHz for individual pulsars (see, for example, \citeauthor{Malofeev1995} (\citeyear{Malofeev1995})) show that the characteristic times of such scintillation are tens of seconds and minutes, and the scale of refractive scintillation are weeks and months (\citeauthor{Smirnova1998}, \citeyear{Smirnova1998}). Moreover, if refractive scintillation change the observed flux density to $20\%$, then diffraction scintillations can be $100\%$. In our monitoring observations, diffraction scintillations are strongly suppressed due to the large width of the frequency channel and only refractive scintillations contribute to the variability.

As for the intrinsic variability of pulsars, there are no detailed statistical studies yet, especially on scales larger than a day. At shorter times from seconds to hours, several dozen pulsars have known intensity variations associated with effects such as mode change, nulling and sub-pulse drift. It is also known that some pulsars have flash phenomena, when they may not be detected for dozens of days, and then for about the same time be similar to ordinary pulsars. This behavior is observed in one of the very first open Pushchino pulsars $J0946+0951$ (\citeauthor{Vitkevich1969} (\citeyear{Vitkevich1969}) and in the pulsar $J1931+2422$ (\citeauthor{Kramer2006}, \citeyear{Kramer2006}) discovered ten years ago. The most exotic behavior is demonstrated by pulsars with rare pulses for tens of minutes of observations, the so-called RRATs (\citeauthor{McLaughlin2006}, \citeyear{McLaughlin2006}) and a completely new type of very rare flares that take about $0.13\%$ of the observation time, which is demonstrated by the pulsar $J0653+8051$, and only the region of the central component in the three-component integral profile (\citeauthor{Malofeev2016}, \citeyear{Malofeev2016}). The search for such pulsars in the power spectra averaged over the entire observation period is not optimal.

A change in the sensitivity of observations may also occur due to an increase in the recording noise track. This increase may be due to interference that could not be completely removed from the processed records, or to a change in the effective area of the antenna due to the weather sensitivity of the LPA LPI. The width of the noise track may also vary due to the effect of confusion of sources scintillating on the interplanetary plasma (\citeauthor{Artyukh1982}, \citeyear{Artyukh1982}), as well as due to an increase in the noise track during daytime at elongations of $20-100^\circ$ associated with known scintillating radio sources (\citeauthor{Chashei2015}, \citeyear{Chashei2015}).

And, finally, the omission of a part of the known pulsars may be associated with a possible flattening of the spectrum at low frequencies, which will especially affect the search for weak pulsars of the ATNF catalog.

The considered sensitivity changes may lead to the fact that of all the records that were selected for the addition of Fourier power spectra, only a small part is suitable for the search. Moreover, for each pulsar under study, it is necessary to summarize its own set of individual power spectra. An increase in sensitivity proportional to the root of the number of accumulated impulses leads to obvious consequences. Let's show this with a simple example. Let one of the pulsars have no variability, its S/N ratio in the accumulated average profile is equal to one per day of observations, the pulsar is observed for 100 days. The second pulsar is flashy and is visible only once in all hundred days, but its S/N ratio on that day was 10 units. Then the S/N ratio when a pulsar is detected by a hundred-day accumulation of Fourier power spectra and in a single recording will be the same. At the same time, if we sum up 100 spectra for the second pulsar, then instead of increasing the S/N ratio, we will get a significant decrease in this value if the desired signal is absent in most of the observations. In Fig.~\ref{fig:tul_fig3_E} shows a graph illustrating the sensitivity changes depending on the width of the noise track in S/N units when accumulated over a different number of days.

Fig.~\ref{fig:tul_fig3_E} shows that a relatively small number of stacked power spectra
having a maximum S/N ratio may be better at searching for pulsars than adding
up all available power spectra.

Thirdly, the use of wide frequency channel bands and poor time resolution degrade the S/N ratio in the source recordings, which are used to obtain the average profile, dynamic spectrum and search for a dispersion measure.

The frequency channel width of $400$~ kHz and the time constant of $100$~ms are too large to organize an optimal search for pulsars at a frequency of $111$~MHz. According to ATNF, in the sample of pulsars with $P>0.4$\,with only $5\%$ pulsars have $W_{0.5} >100$~ms ($W_{0.5}$ is the duration of the half-height integral profile), approximately $13\%$ of pulsars have $W_{0.5}<10$~ms, and $50\%$ of pulsars have $10<W_{0.5}<30$~ms. Consequently, when searching for pulsars in individual recordings, the time constant used is not optimal for $95\%$ of all pulsars and leads to a deterioration in the S/N ratio in the original recording. For more than half of the pulsars, the deterioration will be 2-3 times. At the same time, when pulsars are detected from the averaged power spectra, the deterioration of the S/N in the original recordings does not affect significantly, and with synchronous accumulation during the pulsar search in individual recordings, the deterioration of the S/N will be very noticeable.
Broadening of the pulse inside the frequency channel band will lead to a similar effect of reducing the S/N in the original recordings. We estimate the pulse broadening in the $400$~kHz band for different dispersion measures by the formula:

\begin{equation}\label{DM}
 \Delta t = 4.1488 \cdot 10^6 \cdot (\frac{1}{\nu_2^2}-\frac{1}{\nu_1^2})\cdot DM (ms),
\end{equation}

where $\Delta t$ is the observed pulse broadening, $\nu_1$ and $\nu_2$ are the extreme frequencies of one frequency channel, and DM is a dispersion measure.

For $DM=10~pc/cm^3$, the pulse broadening in the 400~kHz band will be 24~ms, which, taking into account the typical $W_{0.5}$ discussed in the paragraph above, will have practically no effect on the S/N ratio for most of the known pulsars. For $DM=50~pc/cm^3$, the pulse broadening will be 120~ms, which exceeds the accumulation time of one point of the original record and, therefore, starting from this dispersion measure, there will be an additional decrease in the S/N ratio when searching for new pulsars in individual records. For $DM=100$~pc/cm$^3$, the pulse broadening in the frequency channel band will be 240~ms. In fact, the search for such pulsars will be possible if their periods are greater than $0.5$~s. Pulsars with minimal periods, having high of dispersion measures, will look like ideal coaxial rotators. Formula for the sensitivity of pulsar observations (\citeauthor{Dewey1985}, \citeyear{Dewey1985}):

\begin{equation}\label{chuvst}
\Delta S = \frac{\epsilon 2kT}{A_{eff}(\Delta \nu \tau N)^{1/2}} \left( \frac{W}{P-W} \right)^{1/2},
\end{equation}

where $\epsilon=0.7$ is the coefficient for the compensation receiver, $k$ is the Boltzmann constant, $T$ is the temperature of the receiving system, $A_{eff}$ is the effective area of the antenna, $\Delta\nu$ is the reception band, $\tau$ is the time constant, $N$ -- number of accumulated pulses, $P$ -- pulsar period, $W$ -- effective pulse duration. The last multiplier shows that the smaller the pulse width, the higher the fluctuation sensitivity of observations. Consequently, already at $DM=100$ ~pc/cm$^3$, due to the broadening of the pulse within the frequency band, the S/N ratio will additionally decrease by $1.5$ times.

\begin{figure}
\begin{center}
	\includegraphics[width=0.45\textwidth]{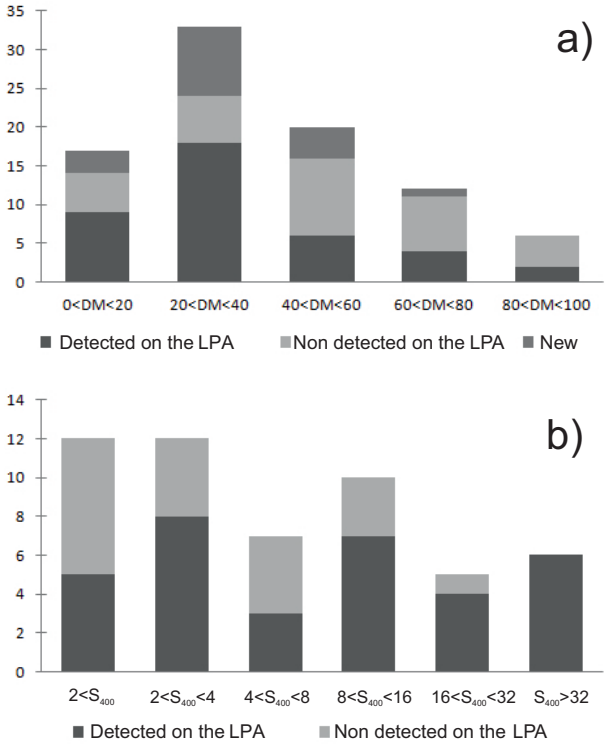}
    \caption{a) Detection of pulsars depending on the DM in monitoring observations; b) Detection of pulsars depending on the flux density at 400 ~MHz}
    \label{fig:tul_fig4_E}
\end{center}
\end{figure}

Thus, the frequency channel bands and time resolution used in the observations can noticeably reduce the S/N ratio already at the dispersion measures of 50-100 $pc/cm^3$. Of course, very strong pulsars having DM>100 $pc/cm^3$ and periods greater than 1 s should also be registered, but not in all cases, taking into account the above arguments.

It is difficult to estimate the total loss of sensitivity during the search for pulsars in individual recordings associated with the frequency channel bandwidth and time resolution used, but it can be expected that the loss of sensitivity $3-5$ times will be typical when searching for pulsars by synchronous accumulation in monitoring data, especially for distant sources.

On~fig.4a shows a histogram that allows you to clearly see how many pulsars are lost during the search. To do this, using ATNF in the region $21^\circ<\delta<42^\circ$, which was used for searching in the early work \cite{Tyulbashev2016}, pulsars with $P>0.4$\,c and with $DM\leq 100 pc/cm^3$ were selected. There were 73 such pulsars in total. The histogram shows that the percentage of lost pulsars increases with the increase in the dispersion measure and the percentage of detected ones decreases. The histogram in Fig.4b was constructed using the ATNF functionality. The histogram shows the number of pulsars detected and not detected during the search for known ATNF, depending on their dispersion measure. It can be seen that the higher the pulsar flux density at 400 MHz, the higher the percentage of detected pulsars.

As shown above, the search for pulsars in the accumulated Fourier power spectra is most advantageous for the preliminary detection of pulsars that do not have variability. In this case, the final S/N ratio depends only on the number of power spectra combined. Note also that the S/N ratio in the first harmonic of the power spectrum will be the maximum for a pure sinusoidal signal and, if we are talking about the search for pulsars, we are talking either about coaxial rotators, in which the energy in the average profile takes up most of the period, or about pulsars with high DM, the pulse of which is smeared, for a significant part of the average profile due to too wide frequency channels used in our observations.

For flare pulsars, for transients, for pulsars with high intrinsic variability, for close pulsars with dispersion measures less than $20 pc/cm^3$, having strong variability due to interstellar scintillation, the search for pulsars by averaging power spectra over the entire observation period will not give much gain, and therefore it is necessary to create a separate the method of searching for such pulsars.

In particular, the "Introduction" mentioned the work \citeauthor{Rodin2015} (\citeyear{Rodin2015}), in which 14 candidates for pulsars detected by individual power spectra were given. In total, 3 months of observations of the LPA LPI were processed, obtained in a six-channel frequency mode and at a time constant of 0.1 s. In this work, a pulsar candidate was considered detected if there are at least two power spectra with matching periodic signals for it. At the same time, the S/N ratio in harmonics corresponding to this periodic signal was at least 5. Our verification of these objects in the averaged power spectra confirmed the existence of two sources. The source $J0926+3018$, having P=1.046 s and $DM=31\pm 2 pc/cm^3$, corresponds, apparently, to the previously open source $J0928+3037$ with P=2.0919 s and $DM=22\pm 2 pc/cm^3$ (\citeauthor{Tyulbashev2016}, \citeyear{Tyulbashev2016}). The source $J1027-0218$, having P=0.539 s and $DM=42\pm 2 pc/cm^3$, is apparently the lower culmination of the pulsar $J2219+4754$ with a period of $P =0.53747$ s and $DM=43.5 pc/cm^3$. It is possible that the remaining 12 candidates have a variability, the reasons for which are discussed in the previous paragraphs, and therefore were not detected in the Fourier averaged power spectra.

\section{Conclusions}

The search for pulsars for periods greater than 0.4 s was carried out on declinations of $-9^\circ<\delta<42^\circ$ and right ascensions $0^h<\alpha<24^h$ on the antenna of the  LPA LPI. 18 new pulsars with periods greater than 0.4 s and with dispersion measures from 5 to $84 pc/cm^3$ were found in the investigated area. A search based on the method of summing Fourier power spectra confirmed the existence of seven pulsars (\citeauthor{Tyulbashev2016}, \citeyear{Tyulbashev2016}), previously discovered during the search, by a direct search of periods and dispersion measures.

\section*{Acknowledgments}

The work was supported by RFBR grants $16-02-00954$ and $15-07-02830$, as well as the program of the Presidium of the Russian Academy of Sciences "Transient and explosive processes in Astrophysics".

\bsp	
\label{lastpage}
\end{document}